\RequirePackage{fix-cm}
\documentclass[twocolumn,epjc3]{svjour3}
\smartqed  
\RequirePackage{graphicx}
 \RequirePackage{mathptmx}      
\RequirePackage[numbers,sort&compress]{natbib}

\usepackage{amsmath}
\usepackage{amssymb}
\usepackage{url}

\newcommand\schbh{Schwarzschild black hole}
\newcommand\schst{Schwarzschild spacetime}

\newcommand\rnst{Reissner-Nordstr\"{o}m spacetime}

\begin{document}

\title{Strong gravitational field time delay for photons coupled to Weyl tensor in a Schwarzschild black hole 
}


\author{Xu Lu \thanksref{addr1,addr2,addr3} \and Feng-Wei Yang \thanksref{addr1,addr2,addr3}
        \and
        Yi Xie\thanksref{addr1,addr2,addr3,e1} 
}

\thankstext{e1}{e-mail: yixie@nju.edu.cn}


\institute{
School of Astronomy and Space Science, Nanjing University, Nanjing 210093, China \label{addr1}
\and
           Shanghai Key Laboratory of Space Navigation and Position Techniques, Shanghai 200030, China \label{addr2}
\and Key Laboratory of Modern Astronomy and Astrophysics, Nanjing University, Ministry of Education, Nanjing 210093, China \label{addr3}
}

\date{Received: date / Accepted: date}

\maketitle

\begin{abstract}
We analyse strong gravitational field time delay for photons coupled to the Weyl tensor in a Schwarzschild black hole. By making use of the method of strong deflection limit, we find that these time delays between relativistic images are significantly affected by polarization directions of such a coupling. A practical problem about determination of the polarization direction by observations is investigated. It is found that if the first and second relativistic images can be resolved, the measurement of time delay can more effectively improve detectability of the polarization direction.
\end{abstract}

\section{Introduction}
\label{sec:intro}

Strong field gravitational lensing by a black hole has generated considerable recent research interest, which is being enhanced by a growing number of efforts to directly image the supermassive black hole at the Galactic center, Sgr A* \cite{Doeleman2008Nature455.78,Fish2011ApJ727.L36,Ricarte2015MNRAS446.1973}. It was demonstrated by Darwin \cite{Darwin1959PRSLSA249.180} that light rays can be significantly deflected in the vicinity of a black hole, resulting relativistic images. Those images are unique features presented by the strong gravitational field of a black hole. By making use of the strong deflection limit (SDL), which requires the light ray very close to the photon sphere \cite{Atkinson1965AJ70.517,Virbhadra2000PRD62.084003,Claudel2001JMP42.818}, an analytical approach can be employed to describe the strong field lensings by the \schst\ \cite{Darwin1959PRSLSA249.180,Luminet1979AA75.228,Ohanian1987AJP55.428,Nemiroff1993AJP61.619,Bozza2001GRG33.1535,Eiroa2004PRD69.063004}, by the \rnst\ \cite{Eiroa2002PRD66.024010,Eiroa2004PRD69.063004}, and by a static and spherically symmetric spacetime \cite{Bozza2002PRD66.103001}. With the SDL method, astronomical observables can be easily obtained, including the positional separations, the brightness differences and time delays among the relativistic images. The strong field lensings by static and spherically symmetric spacetimes were also considered in various contexts with different methods \cite{Virbhadra1998AA337.1,Virbhadra2000PRD62.084003,Bhadra2003PRD67.103009,Perlick2004PRD69.064017,Whisker2005PRD71.064004,Majumdar2005IJMPD14.1095,Eiroa2005PRD71.083010,Keeton2006PRD73.044024,Eiroa2006PRD73.043002,Amore2006PRD73.083004,Amore2006PRD74.083004,Amore2007PRD75.083005,Iyer2007GRG39.1563,Mukherjee2007GRG39.583,Bozza2008PRD78.103005,Pal2008CQG25.045003,Virbhadra2008PRD77.124014,Virbhadra2009PRD79.083004,Chen2009PRD80.024036,Liu2010PRD81.124017,Bin-Nun2010PRD81.123011,Bin-Nun2010PRD82.064009,Eiroa2011CQG28.085008,Eiroa2012PRD86.083009,Eiroa2013PRD88.103007,Gyulchev2013PRD87.063005,Eiroa2014EPJC74.3171,Wei2015EPJC75.253,Wei2015EPJC75.331,Man2015PRD92.024004,Sotani2015PRD92.044052}. Strong field lensings of rotating black holes were discussed \cite{Bozza2003PRD67.103006,Vazquez2004NCimB119.489,Bozza2005PRD72.083003,Bozza2006PRD74.063001,Bozza2007PRD76.083008,Gyulchev2007PRD75.023006,Bozza2008PRD78.063014,Chen2010CQG27.225006,Chen2011PRD83.124019,Kraniotis2011CQG28.085021,Chen2012PRD85.124029,Kraniotis2014GRG46.1818,Ji2014JHEP03.089,Cunha2015PRL115.211102}. Relativistic images might be able to provide new hints for the possible existence of naked singularities \cite{Virbhadra2002PRD65.103004,Virbhadra2008PRD77.124014,Sahu2012PRD86.063010,Sahu2013PRD88.103002,Gyulchev2008PRD78.083004} as well as wormholes \cite{Kuhfittig2014EPJC74.2818,Kuhfittig2015arXiv1501.06085,Nandi2006PRD74.024020,Tsukamoto2012PRD86.104062}. Reviews of strong gravitational field lensing can be found in \cite{Bozza2010GRG42.2269,Eiroa2012arXiv1212.4535}.

The underlying reason of gravitational lensing is the interaction between electromagnetic and gravitational fields. Beyond the standard Einstein-Maxwell theory, the authors of \cite{Drummond1980PRD22.343} investigated the local propagation of photons after considering the effects of one-loop vacuum polarization on the photon effective action for quantum electrodynamics. It was found \cite{Drummond1980PRD22.343} that the properties of light propagation could be changed by tidal gravitational forces introduced by these quantum corrections, and a photon might be able to travel ``faster than light'' in some cases. These ``superluminal'' photons were also found in various gravitational contexts \cite{Mankiewicz1989PRD40.2134,Khriplovich1995PLB346.251,Daniels1994NPB425.634,Shore1996NPB460.379,Daniels1996PLB367.75,Mohanty1998NPB526.501,Cho1997PRD56.6416,Cai1998NPB524.639,Shore2001NPB605.455,Shore2002NPB633.271}. However, strength of the effects is immeasurably small because its coupling constant is inversely proportional to $\lambda^2_e$, where $\lambda_e$ is the Compton wavelength of an electron \cite{Drummond1980PRD22.343}. Motivated by some physical circumstances, extended theoretical models without this limit on the coupling were investigated for primordial magnetic fields \cite{Turner1988PRD37.2743,Mazzitelli1995PRD52.6694,Lambiase2004PRD70.063502,Raya2006PhLB638.314,Campanelli2008PRD77.123002,Bamba2008JCAP04.024} and new coupling between gravity and photons \cite{Ni1977PRL38.301,Ni1984PMFC.647,Lafrance1995PRD51.2584,Novello1996CQG13.1089,Hehl2001LNP562.479,Itin2003PRD68.127701,Solanki2004PRD69.062001,Preuss2004PRD70.067101,Balakin2005CQG22.1867,Balakin2008PRD77.084013,Dereli2011EPJC71.1589}.    

Another way to couple the electromagnetic and gravitational fields can be realized through Weyl tenser in the effective action, which has been widely investigated in the holographic conductivity and superconductor \cite{Ritz2009PRD79.066003,Ma2011PLB704.604,Wu2011PLB697.153,Momeni2011MPLA26.2889,Momeni2012IJMPA27.1250128,Momeni2012EPL97.61001,Roychowdhury2012PRD86.106009,Zhao2013PLB719.440,Jing2016AoP367.219} and dynamical evolution of electromagnetic field in the black hole spacetime \cite{Chen2013PRD88.064058,Chen2014PRD90.124059,Chen2014PRD89.104014,Liao2014PLB728.457}. The authors of \cite{Chen2015JCAP10.002} studied the strong field gravitational lensing for the photons coupled to Weyl tensor in the \schbh, and obtained the strong deflection angle, angular separations and brightness difference between resulting relativistic images. Complementary to these observables, time delays between these images are important observables as well. They can be used to measure the distance of the black hole \cite{Bozza2004GRG36.435} and can also be probe of cosmic censorship \cite{Sahu2013PRD88.103002} and the Gauss-Bonnet correction \cite{Man2014JCAP11.025}. 

In this work, as an extension of the previous work \cite{Chen2015JCAP10.002}, we will focus on the strong field time delays between relativistic images when such a coupling of the Wely tensor is taken into account. In Sect. \ref{sec:model}, the effective metric for the Weyl tensor coupled photons is briefly reviewed for completeness. The strong field time delay is investigated in Sect. \ref{sec:td}. The observables of such delays between relativistic images are discussed in Sect. \ref{sec:obs} by taking Sgr A* as an example. A practical problem about determination of the polarization direction for the coupling of the Wely tensor by observations is also investigated. Finally, in Sect. \ref{sec:con}, we summarize our results.

\section{Effective metric for Weyl tensor coupled photons}

\label{sec:model}

The effective metric for the Weyl tensor coupled photons will be briefly reviewed for completeness in this section, which only covers necessary information for our following work. More details can be found in \cite{Drummond1980PRD22.343,Chen2015JCAP10.002,Jing2016AoP367.219}. We consider the electromagnetic field coupled to the Weyl tensor in the curved spacetime as (in the units $G=c=1$) \cite{Ritz2009PRD79.066003}
\begin{equation}
  \label{eq:action}
  S  = \int\sqrt{-g}\mathrm{d}^4x\bigg[\frac{R}{16\pi}-\frac{1}{4}\bigg(F_{\mu\nu}F^{\mu\nu}-4\alpha C^{\mu\nu\rho\sigma}F_{\mu\nu}F_{\rho\sigma}\bigg)\bigg],\nonumber\\
\end{equation}
where the $4$-dimensional Weyl tensor $C_{\mu\nu\rho\sigma}$ of the spacetime metric $g_{\mu\nu}$ is defined as
\begin{equation}
  \label{}
  C_{\mu\nu\rho\sigma} = R_{\mu\nu\rho\sigma}-(g_{\mu[\rho}R_{\sigma]\nu}-g_{\nu[\rho}R_{\sigma]\mu}) +\frac{1}{3}Rg_{\mu[\rho}g_{\sigma]\nu}.
\end{equation}
The square brackets around indices are used to denote the antisymmetric parts. $F_{\mu\nu}\equiv A_{\nu;\mu}-A_{\mu;\nu}$ is the electromagnetic tensor. $\alpha$ is the coupling constant with dimension of [Length]$^2$. Variation of the above action with respect to $A_{\mu}$ gives the Weyl tensor corrected Maxwell equation as
\begin{equation}
\label{eq:eomF}
\bigg(F_{\mu\nu}-4C^{\mu\nu\rho\sigma}F_{\rho\sigma}\bigg)_{;\mu}=0.
\end{equation}
With the geometric optics approximation that $\lambda_e<\lambda<L$, where $\lambda$ is the wavelength of the photon and $L$ is the typical curvature length scale, we can set \cite{Drummond1980PRD22.343}
\begin{equation}
  \label{eq:Fuvf}
  F_{\mu\nu}=f_{\mu\nu}\exp(i\theta),
\end{equation}
and regard $f_{\mu\nu}$ are slowly varying compared with $\theta$. Putting $k_{\mu}=\theta_{;\mu}$ and using the electromagnetic Bianchi identity, we can have \cite{Drummond1980PRD22.343}
\begin{equation}
  \label{}
  k_{\rho}f_{\mu\nu}+k_{\mu}f_{\nu\rho}+k_{\nu}f_{\rho\mu}=0,
\end{equation}
which leads to 
\begin{equation}
  \label{eq:fuv}
  f_{\mu\nu} = k_{\mu}a_{\nu}-k_{\nu}a_{\mu}.
\end{equation}
Here, $a_{\mu}$ is the polarization vector which satisfies $k_{\mu}a^{\mu}=0$. After rewriting Eq. \eqref{eq:eomF} with Eqs. \eqref{eq:Fuvf} and \eqref{eq:fuv}, we can obtain the equations of motion for a photon which couples to the Weyl tensor as \cite{Chen2015JCAP10.002,Jing2016AoP367.219}
\begin{equation}
  \label{}
  k_{\mu}k^{\mu}a^{\nu}+8\alpha C^{\mu\nu\rho\sigma}k_{\sigma}k_{\mu}a_{\rho}=0.
\end{equation}

If considering a 4-dimensional static and spherically symmetric spacetime as the background for the photon's propagation, i.e.,
\begin{equation}
  \label{eq:metric}
  \mathrm{d}s^2=-f(r)\mathrm{d}t^2+\frac{\mathrm{d}r^2}{f(r)}+r^2(\mathrm{d}\theta^2+\sin^2{\theta}\mathrm{d}\varphi^2),
\end{equation}
we can introduce the orthonormal tetrad \cite{Drummond1980PRD22.343}
\begin{equation}
  \label{}
  e^{a}_{\mu}=\bigg(\sqrt{f},\frac{1}{\sqrt{f}},r,r\sin\theta\bigg)
\end{equation}
and the bivectors \cite{Drummond1980PRD22.343}
\begin{equation}
  \label{}
  U^{ab}_{\mu\nu} = e^a_{\mu}e^b_{\nu}-e^{a}_{\nu}e^{b}_{\mu}.
\end{equation}
In order to simplify the equations of motion for the coupled photon, three independent vectors can be defined as \cite{Drummond1980PRD22.343}
\begin{equation}
  \label{}
  l_{\nu} = k^{\mu}U^{01}_{\mu\nu},\quad n_{\nu}=k^{\mu}U^{02}_{\mu\nu},\quad m_{\nu} = k^{\mu}U^{23}_{\mu\nu},
\end{equation}
which are all orthogonal to the vector $k_{\nu}$. Therefore, the light-cone conditions can be found out as \cite{Chen2015JCAP10.002,Jing2016AoP367.219}
\begin{equation}
  \label{}
  (g_{00}k^0k^0+g_{11}k^1k^1)+W(g_{22}k^2k^2+g_{33}k^3k^3)=0,
\end{equation}
where $W$ depends on polarization of the photon. When the polarization is along the direction of $l_{\mu}$ (PPL), it is
\begin{equation}
  \label{}
  W(r)=\frac{r^3-8\alpha M}{r^3+16\alpha M}; 
\end{equation}
when the polarization is along $m_{\mu}$ (PPM), then it is
\begin{equation}
  \label{}
  W(r)=\frac{r^3+16\alpha M}{r^3-8\alpha M}. 
\end{equation}
Although these light-cone conditions demonstrate that the coupled photon no longer travels as null geodesic worldline in the \schst, an effective metric can be constructed to make it null geodesic \cite{Breton2002CQG19.601}. When we take $2M$ as the measure of distances and set it to unity, the effective metric can be written as \cite{Chen2015JCAP10.002,Jing2016AoP367.219}
\begin{equation}
  \label{eq:emetric}
  \mathrm{d}s^2=-A(x)\mathrm{d}t^2+B(x)\mathrm{d}x^2+C(x)(\mathrm{d}\theta^2+\sin^2{\theta}\mathrm{d}\varphi^2),
\end{equation}
where the functions are
\begin{eqnarray}
  \label{}
  A(x) & = & B(x)^{-1}=1-\frac{1}{x},\\
  C(x) & = & x^2 W(x),\\
  W(x) & = & W_{\mathrm{PPL}} = \frac{x^3-4\alpha}{x^3+8\alpha} \qquad \mathrm{for\quad PPL},\\
  W(x) & = & W_{\mathrm{PPM}} = \frac{x^3+8\alpha}{x^3-4\alpha} \qquad \mathrm{for\quad PPM}.
\end{eqnarray}
Following the assumption of \cite{Chen2015JCAP10.002}, we also consider that the observer, the source and the path of the photon are all located in the plane of $\theta=90^{\circ}$ in the \schst, while the observer and the source are very much far away from the black hole. It was also found \cite{Chen2015JCAP10.002} that, in order to ensure a photon always stay outside the event horizon, $\alpha$ must satisfy the conditions: $4\alpha<1$ for PPL and $8\alpha>-1$ for PPM.

\section{Time delays between relativistic images}

\label{sec:td}

In this section, we will follow the approach proposed in \cite{Bozza2004GRG36.435} and calculate the time delays between relativistic images of the Weyl tensor coupled photons. For the time component of a null geodesic in the spacetime \eqref{eq:emetric}, we can have \cite{Bozza2004GRG36.435}
\begin{equation}
	\dfrac{\mathrm{d}t}{\mathrm{d}x}=\tilde{P}_1(x,x_0)P_2(x,x_0),
\end{equation}
where the two functions are
\begin{eqnarray}
  \label{}
  \tilde{P}_1(x,x_0) & = & \sqrt{\frac{BA_0}{A}},\\
  P_2(x,x_0) & = & \frac{1}{\sqrt{A_0-A\frac{C_0}{C}}}.
\end{eqnarray}
The subscript ``$0$'' of a quantity means its value at $x=x_0$, where $x_0$ is the closest distance of the photon to the black hole. The time taken by a photon from the source to the observer can be decomposed into three parts \cite{Bozza2004GRG36.435}
\begin{equation}
  \label{appeq:T3parts}
  T =  \tilde{T}(x_0) - \int^{\infty}_{D_{\mathrm{OL}}}\bigg|\frac{\mathrm{d}t}{\mathrm{d}x}\bigg|\mathrm{d}x - \int^{\infty}_{D_{\mathrm{LS}}}\bigg|\frac{\mathrm{d}t}{\mathrm{d}x}\bigg|\mathrm{d}x,
\end{equation}
where $D_{\mathrm{OL}}$ and $D_{\mathrm{LS}}$ are the distances of observer-lens and lens-source, and $\tilde{T}(x_0)$ is defined as \cite{Bozza2004GRG36.435}
\begin{equation}
\label{eq:dftime}
\tilde{T}(x_0)=\int^{\infty}_{x_0}\bigg|\dfrac{\mathrm{d}t}{\mathrm{d}x}\bigg|\mathrm{d}x = \int^{\infty}_{x_0}{2\sqrt{B(x)C(x)A_0}\over{A(x)\sqrt{C_0}\sqrt{{C(x)\over{C_0}}{A_0\over{A(x)}}-1}}}\mathrm{d}x.
\end{equation}
It is assumed that the observer and the source are far from the lens, the time delay between two relativistic images 1 and 2 can be given as \cite{Bozza2004GRG36.435}
\begin{eqnarray}
  T_1-T_2&=&2\int^\infty_{x_{0,1}}\left|\dfrac{\mathrm{d}t}{\mathrm{d}x}(x,x_{0,1})\right|\mathrm{d}x-2\int^\infty_{x_{0,2}}\left|\dfrac{\mathrm{d}t}{\mathrm{d}x}(x,x_{0,2})\right|\mathrm{d}x,\\
&=&\tilde{T}(x_{0,1})-\tilde{T}(x_{0,2})+2\int^{x_{0,2}}_{x_{0,1}}\dfrac{\tilde{P}_1(x,x_{0,1})}{\sqrt{A_{0,1}}}\mathrm{d}x,
\end{eqnarray}
where the subscript ``${,i}$'' ($i=1,2$) of a quantity is its value of the $i$-th relativistic image.

With the technique of SDL \cite{Bozza2002PRD66.103001}, the integral of $\tilde{T}(x_0)$ can be rewritten as \cite{Bozza2004GRG36.435}
\begin{equation}
  \label{}
  \tilde{T}(x_{0})=\int^1_0\tilde{R}(z,x_{0})f(z,x_{0})\mathrm{d}z,
\end{equation}
where the quantity $z$ is defined as
\begin{equation}
	z=\frac{A-A_0}{1-A_0},
\end{equation}
and the two functions are
\begin{eqnarray}
f(z,x_0) & = & P_2(x,x_0),\\
\tilde{R}(z,x_{0}) & = & 2\dfrac{1-A_{0}}{A'(x)}\tilde{P}_1(x,x_{0})\left(1-\dfrac{1}{\sqrt{A_{0}}f(z,x_{0})}\right).
\end{eqnarray}
The prime means partial derivative against $x$. In order to find $\tilde{T}(x_{0})$ in the SDL, we define the radius of the photon sphere $x_m$ as \cite{Virbhadra2000PRD62.084003,Claudel2001JMP42.818}
\begin{equation}
\label{eq:phosph}
\dfrac{C'(x)}{C(x)}=\dfrac{A'(x)}{A(x)},
\end{equation}
which leads to
\begin{equation}
  \label{}
  (x^3+8\alpha)(x^3-4\alpha)(2x-3)\pm36\alpha x^3(x-1)=0.
\end{equation}
In the last term of the above equation, the plus sign originates from the case of PPL and the minus one comes from the PPM one. The biggest root of this equation can be taken as $x_m$, which can be solved numerically. We can obtain $\tilde{T}(x_{0})$ in the SDL at $x_0\sim x_m$ and transform the variable $x_0$ to the impact parameter $u$, which is given by \cite{Weinberg1972Book,Virbhadra2000PRD62.084003}
\begin{equation}
u=\sqrt{\dfrac{C_0}{A_0}}
\end{equation}
and its value at the photon sphere $x_0=x_m$ is $u_m$. Finally, we can have \cite{Bozza2004GRG36.435}
\begin{equation}
  \label{}
  \tilde{T}(u)=-\tilde{a}\ln{\left(\dfrac{u}{u_m}-1\right)}+\tilde{b}+\mathcal{O}(u-u_m),
\end{equation}
where $\tilde{a}$ and $\tilde{b}$ are the coefficients. Fortunately, although the exact expressions of these two coefficients are complicated and can be found in \cite{Bozza2004GRG36.435}, they are not directly needed in the following calculation.

If we assume the source, the lens and the observer are aligned almost in a line and use an approximated relation that \cite{Bozza2004GRG36.435}
\begin{equation}
	 \int^{x_{0,2}}_{x_{0,1}}\dfrac{\tilde{P}_1(x,x_{0,1})}{\sqrt{A_{0,1}}}\mathrm{d}x\approx\sqrt{\dfrac{B_m}{A_m}}\ (x_{0,2}-x_{0,1})
\end{equation}
where the subscript $m$ of a quantity means its value at $x=x_m$, we can have the time delay between a $n$-loop and a $m$-loop relativistic images as \cite{Bozza2004GRG36.435}
\begin{equation}
\label{DTnm}
\Delta T_{n,m} = \Delta T_{n,m}^{\ 0}+\Delta T_{n,m}^{\ 1},
\end{equation}
where the leading term $\Delta T_{n,m}^{\ 0}$ and its correction $\Delta T_{n,m}^{\ 1}$ are
\begin{eqnarray}
\Delta T_{n,m}^{\ 0} & = & 2 \pi (n-m) u_m,\\
\Delta T_{n,m}^{\ 1} & = & 2\sqrt{\frac{B_m}{A_m}} \sqrt{\dfrac{u_m}{\hat c}}\exp\left({\frac{\bar{b}}{2\bar{a}}}\right)\nonumber\\
  & & \quad\times \left[\exp\left({-\frac{m\pi}{\bar{a}}}\right)-\exp\left({-\frac{n\pi}{\bar{a}}}\right)\right].
\end{eqnarray}
Here, we used a relation for a spherically symmetric metric that is $\tilde{a} = \bar{a}u_m$ \cite{Bozza2004GRG36.435}. The quantities $\bar{a}$, $\bar{b}$ and $\hat{c}$ are \cite{Bozza2002PRD66.103001}
\begin{eqnarray}
 \bar{a} & = & \frac{R(0,x_m)}{2\sqrt{\beta_m}},\\
 \bar{b} & = & -\pi+b_R+\bar{a}\ln{2\beta_m\over{A_m}},\\
\hat{c} & = &  \beta_m\sqrt{\dfrac{A_m}{C_m^3}}\dfrac{{C_m'}{}^2}{2(1-A_m)^2},
\end{eqnarray}
where
\begin{eqnarray}
  \label{}
  R(0,x_m)&=&2\frac{(1-A_m)\sqrt{B_mA_m}}{A'_m\sqrt{C_m}},\\
  \beta_m&=&\frac{C_m(1-A_m)^2 (A_m C_m''-C_m A_m'')}{2A_m^2C_m'^2},\\
  b_R & = & \int^1_0 \left[R(z,x_m)f(z,x_m)-\frac{R(0,x_m)}{\sqrt{\beta_mz}}\right]\mathrm{d}z.
\end{eqnarray}
When these quantities are calculated for a given $\alpha$, the time delays between relativistic images can be worked out.

\section{Observables and detectability of polarization direction}

\label{sec:obs}

In this section, we will take the supermassive black hole Sgr A* with mass $M_{\bullet}=4.31\times10^6$ $M_\odot$ and distance $D_{\mathrm{OL}}= D_{\bullet}=8.33$ kpc \cite{Gillesse2009ApJ707.L114} as an example to evaluate the observables of the strong field time delays.

\begin{figure}[tbp]
\centering
\includegraphics[width=.5\textwidth]{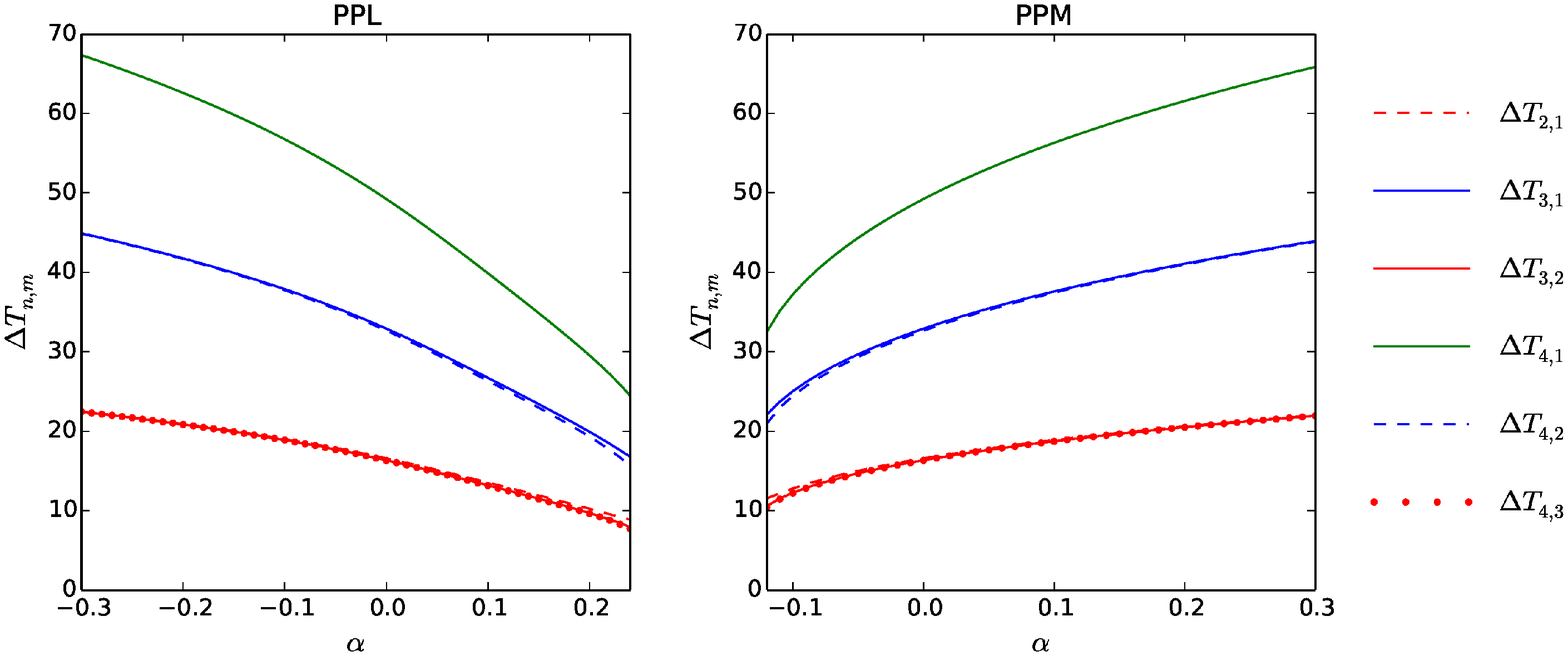}
\caption{\label{fig:DT} Time delays between different relativistic images for Sgr A*.  $\Delta T_{n,m}$ is represented in the unit of $2GM_{\bullet}/c^{3}\approx42.45$ s. The left panel shows the time delays for the case of PPL ($4\alpha<1$) and the right one is for PPM ($8\alpha>-1$).}
\end{figure}

Figure \ref{fig:DT} show the estimated time delays between different relativistic images of the photons coupled to the Weyl tensor in the \schst. $\Delta T_{n,m}$ is represented in the unit of $2GM_{\bullet}/c^{3}\approx42.45$ s. Based on the values of $m$ and $n$, we consider six cases: ($n=2$, $m=1$), ($n=3$, $m=2$), ($n=4$, $m=3$), ($n=3$, $m=1$), ($n=4$, $m=2$) and ($n=4$, $m=1$). It is clearly shown that the time delays grow with the increment of the difference between $n$ and $m$ and the time delays are almost the same for those with the same value of $n-m$. The left panel shows the time delays for the case of PPL and the right one is for PPM. It is distinct that $\Delta T_{n,m}$ responses to the increase of $\alpha$ inversely for the cases of PPL and PPM. When $\alpha=0$, the values of $\Delta T_{n,m}$ reduce to the ones of the \schst. 

\begin{figure}[tbp]
\centering
\includegraphics[width=0.5\textwidth]{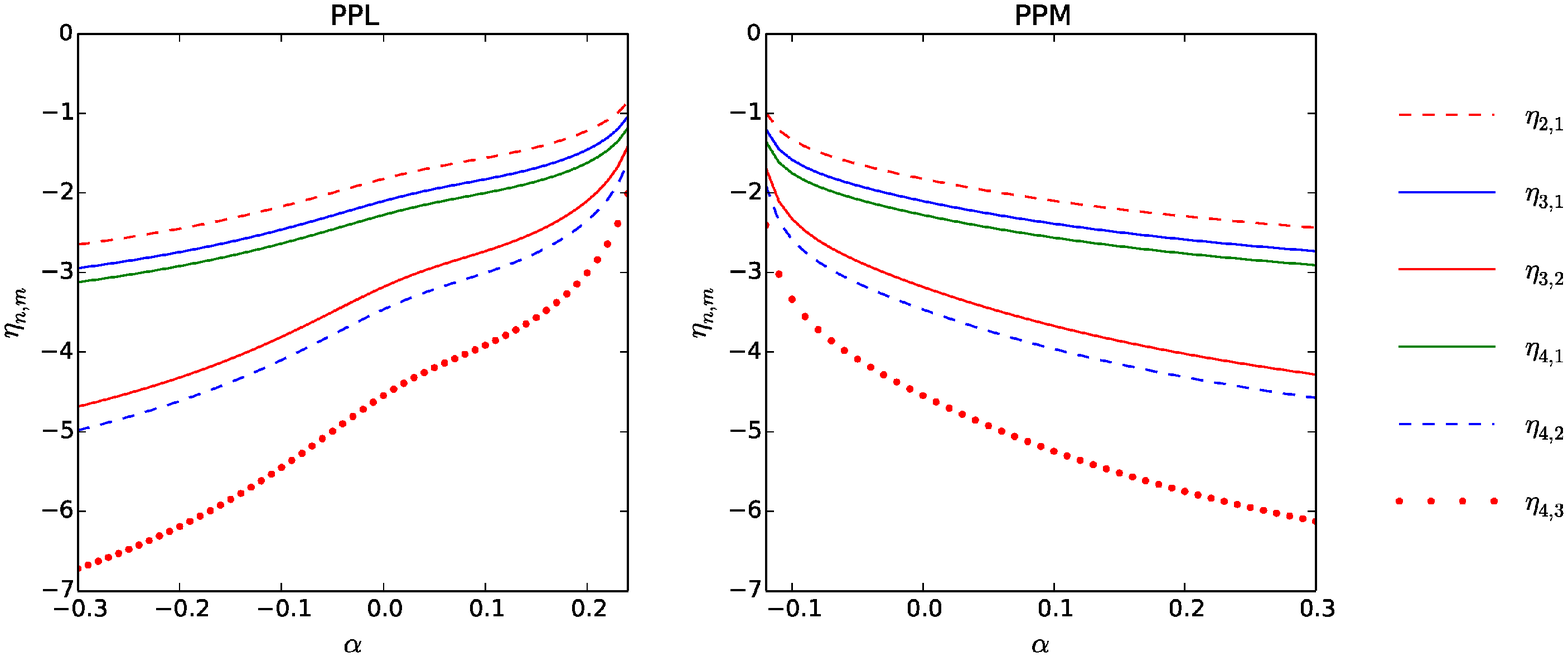}
\caption{\label{fig:eta} The logarithmic contributions of the correction term $\Delta T_{n,m}^{\ 1}$ for the whole time delay. The left panel shows the contributions for the case of PPL ($4\alpha<1$) and the right one is for PPM ($8\alpha>-1$).}
\end{figure}

In order to show the contributions of the correction term $\Delta T_{n,m}^{\ 1}$ for the whole time delay, we define an indicator as
\begin{equation}
  \label{}
  \eta_{n,m} \equiv \log_{10}\bigg(\frac{\Delta T_{n,m}^{\ 1}}{\Delta T_{n,m}}\bigg).
\end{equation}
Figure \ref{fig:eta} clearly show the contributions of $\Delta T_{n,m}^{\ 1}$ are much smaller than those $\Delta T_{n,m}^{\ 0}$ by 1 to 6 orders of magnitude. For a given $\alpha$ and the same loop difference $n-m$, $\eta_{n,m}$ decreases rapidly when $n$ or $m$ increases. $\eta_{n,m}$ also shows responses to the increase of $\alpha$ inversely for the cases of PPL (left panel) and PPM (right panel). 

\begin{figure}[tbp]
\centering
\includegraphics[width=0.48\textwidth]{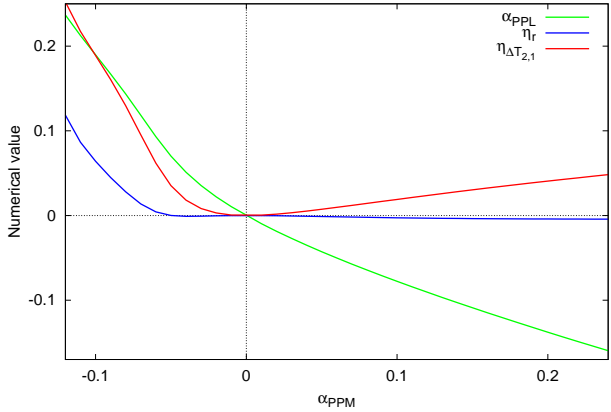}
\caption{\label{fig:poldir} The green curve shows $\alpha_{\mathrm{PPL}}$ against $\alpha_{\mathrm{PPM}}$, both of which satisfy Eq. \eqref{eq:sL=sM} and can generate an identical angular separation of the first and second relativistic images. The contrast indicators $\eta_{r}$ and $\eta_{\Delta T_{2,1}}$ are respectively represented the blue and red curves, which demonstrate $\eta_{\Delta T_{2,1}}$ is considerably larger than $\eta_{r}$.  }
\end{figure}

When the first and second relativistic images can be resolved and their angular separation $s$, brightness difference $r$ and time delay $\Delta T_{2,1}$ are measured, an interesting question will be raised whether the directions of polarization PPL and PPM can be detected and determined, which has not been discussed before. The expressions of $s$ and $r$ for these two polarizations can be found in \cite{Chen2015JCAP10.002}, while $\Delta T_{2,1}$ is given by Eq. \eqref{DTnm} in the present work. Here, we assume the uncertainty of measured $s$ is much smaller than those of the others, because the accuracies of $r$ and $\Delta T_{2,1}$ depend on how clearly these two images are separated. With different values of coupling constant $\alpha_{\mathrm{PPL}}$ and $\alpha_{\mathrm{PPM}}$, both PPL and PPM cases can generate an identical $s$. In Fig. \ref{fig:poldir}, the green curve shows the value of $\alpha_{\mathrm{PPL}}$ against $\alpha_{\mathrm{PPM}}$, and each point $(\alpha_{\mathrm{PPM}}, \alpha_{\mathrm{PPL}})$ on the curve satisfies the relation as
\begin{equation}
  \label{eq:sL=sM}
  s[W_{\mathrm{PPL}}(\alpha_{\mathrm{PPL}})] = s[W_{\mathrm{PPM}}(\alpha_{\mathrm{PPM}})].
\end{equation}
In order to indicate the contrasts of $r$ and $\Delta T_{2,1}$ given by these two polarization directions, we define two quantities as
\begin{eqnarray}
  \label{}
  \eta_r & \equiv & 2\frac{r_{\mathrm{PPM}}-r_{\mathrm{PPL}}}{r_{\mathrm{PPM}}+r_{\mathrm{PPL}}},\\
  \eta_{\Delta T_{2,1}} & \equiv & 2\frac{\Delta T^{\mathrm{PPM}}_{2,1}-\Delta T^{\mathrm{PPL}}_{2,1}}{\Delta T^{\mathrm{PPM}}_{2,1}+\Delta T^{\mathrm{PPL}}_{2,1}},
\end{eqnarray}
where ``PPL''/``PPM'' means the value is obtained according to the case of PPL/PPM, and $\alpha_{\mathrm{PPL}}$ and $\alpha_{\mathrm{PPM}}$ in the above quantities must satisfy Eq. \eqref{eq:sL=sM}. For the brightness difference $r$, $\eta_r$ give the discrepancy between $r_{\mathrm{PPL}}$ and $r_{\mathrm{PPM}}$ which are respectively predicted by PPL and PPM, and this discrepancy is normalized by the average value of them. The contrast indicator $\eta_{\Delta T_{2,1}}$ represents the normalized discrepancy of predicted time delays. Based on the definitions, a larger indicator means two observables associated with it can be more easily distinguished so that the polarization direction can be more clearly determined or ruled out. In Fig. \ref{fig:poldir}, $\eta_r$ against $\alpha_{\mathrm{PPM}}$ is shown by the blue curve, and $\eta_{\Delta T_{2,1}}$ is represented by the red one. At the point of $\alpha_{\mathrm{PPM}}=0$, both $\eta_{r}$ and $\eta_{\Delta T_{2,1}}$ reduce to 0, which is a natural outcome because the coupling of photons to the Wely tensor vanishes.  It is found that $\eta_{r}$ can reach about $0.1$ only when $\alpha_{\mathrm{PPM}}<-0.1$; and it is almost equal to $0$ on the rest and larger part of the domain of $\alpha_{\mathrm{PPM}}$. In contrast to it , $\eta_{\Delta T_{2,1}}$ is considerably bigger than $\eta_{r}$. It can reach about 0.2 when $\alpha_{\mathrm{PPM}}<-0.1$; and when $\alpha_{\mathrm{PPM}}>0.15$, it can reach about 0.05, which is larger than $\eta_{r}$ by a factor of 10. It suggests when the first and second relativistic images are resolved, the measurement of time delay between them can more effectively improve the detectability of the polarization direction than the measurement of bright difference does.

\section{Conclusions and discussion}

\label{sec:con}

In this work, as an extension of the previous work \cite{Chen2015JCAP10.002}, we analyse the strong field time delay for the photons coupled to the Weyl tensor in the \schbh. By making use of the method of SDL \cite{Bozza2002PRD66.103001,Bozza2004GRG36.435}, we calculate the time delays between relativistic images. We find that these time delays are affected by both the coupling constant $\alpha$ and the direction of polarization, and they show responses to the increase of $\alpha$ inversely for the cases of PPL (see left panels of Figs. \ref{fig:DT} and \ref{fig:eta}) and PPM (see right panels of Figs. \ref{fig:DT} and \ref{fig:eta}).

Although it would be very challenging, if the outermost two relativistic images are resolved and their time signals are observed, such a measurement of time delay can verify the observations of their angular separation and brightness difference. Based on the results of \cite{Chen2015JCAP10.002} and ours, it is found that the observations of strong field gravitational lensing, including angular separations, brightness differences and time delays, can possibly detect such a coupling between the photons and the Weyl tensor by comparing the observables with those of the \schst. Furthermore, we also find that, after the resolution of the first and second images, the measurement of time delay can more effectively improve the detectability of the polarization direction than the measurement of bright difference does (see Fig. \ref{fig:poldir}).

\begin{acknowledgements}
This work is funded by the National Natural Science Foundation of China (Grants No. 11573015 and No. J1210039) and the Opening Project of Shanghai Key Laboratory of Space Navigation and Position Techniques (Grant No.14DZ2276100).
\end{acknowledgements}

\bibliographystyle{spphys}       
\bibliography{Refs20160609}   

%
%

\end{document}